\documentclass[12pt, a4paper]{article}

\usepackage{amsmath,amssymb,amsfonts}
\usepackage{graphicx}
\usepackage[colorlinks,hypertexnames=false]{hyperref}
\usepackage{cite}

\begin{document}

\title{Dynamical friction in ultralight dark matter:\\  Plummer sphere perspective}
\author{V.M.~Gorkavenko${}^{1,}$\thanks{Corresponding author. \textit{Email address:} \textbf{gorkavol@knu.ua} (Volodymyr Gorkavenko)},
A.I. Yakimenko${}^{1,3,4}$,
A.O. Zaporozhchenko${}^1$,
E.V.~Gorbar${}^{1,2}$
\\
${}^1$ \it \small Faculty of Physics, Taras Shevchenko National University of Kyiv,\\
\it \small 64, Volodymyrs'ka str., Kyiv 01601, Ukraine\\
${}^2$ \it \small Bogolyubov Institute for Theoretical Physics, National Academy of Sciences of Ukraine,\\
\it \small 14-b, Metrolohichna str., Kyiv 03143, Ukraine\\
${}^3$ \it \small Dipartimento di Fisica e Astronomia ’Galileo Galilei’, Universit$\acute a$ di Padova,\\ \it \small via Marzolo 8, 35131 Padova, Italy\\
${}^4$ \it \small Istituto Nazionale di Fisica Nucleare, Sezione di Padova, via Marzolo 8, 35131 Padova, Italy
}
\date{}

\maketitle
\setcounter{equation}{0}
\setcounter{page}{1}%

\begin{abstract}
In models of dark matter composed of feebly interacting ultralight bosons in the state of Bose-Einstein condensate, the dynamical friction force acting on circularly moving globular clusters modelled as Plummer spheres is determined. Analytic expressions for both radial and tangential components of the dynamical friction force are given. We reveal that the dynamical friction force for the Plummer sphere deviates from that for a point probe of the same mass  for 
significantly large ratio of the Plummer sphere radius to its orbital radius  
as well as for large values of the Mach number.

\end{abstract}

\section{Introduction}

Dynamical friction of moving objects in a galactic environment is a time honed and much studied subject that originated from the seminal work by Chandrasekhar \cite{Chandrasekhar} on dynamical friction force acting on a moving star due to the fluctuating gravitational force of neighbouring stars. Later the dynamical friction for motion in a gaseous medium was considered in \cite{Bondi,Dokuchaev,Ruderman,Rephaeli,Ostriker}.

For point probes moving on circular orbits in a gaseous medium, the dynamical friction force was numerically studied in \cite{Sanchez, Kim:2007,Kim:2008}. An analytical solution to the
dynamical friction force acting on a circularly-moving perturber in a gaseous environment was derived in \cite{Desjacques_2022}. It was found that while the radial force is dominant at large Mach numbers, it is negligible for subsonic motion.

Models of ultra-light dark matter (ULDM) with a mass of DM particles  {in the range $10^{-23}-10^{-21}$ eV have quite interesting phenomenology and are actively studied in the literature (for a review, see \cite{Chavanis:2015zua,Niemeyer:2019aqm,Hui:2021tkt,Ferreira}).} They could successfully reproduce the large-scale structure of the Universe, which is successfully explained in cold dark matter (CDM) models, and are free of some problems which CDM models encounter at the galactic scale. The ULDM models are characterized by the presence of a core in the form of the Bose-Einstein condensate (BEC) of ultra-light bosons. The distribution of the DM in galaxies, including ULDM, is reviewed in \cite{Salucci:2018hqu}.

In the case of ULDM models, the dynamical friction force for moving stars in fuzzy dark matter (ultralight bosonic dark matter without self-interaction) was studied in \cite{Hui:2016ltb,Lancaster_2020,Wang,Boey:2024dks}. According to the analysis of the observation of the mass distribution of galaxy M87, one can conclude that fuzzy dark matter cannot reproduce the observed cores in large galaxies \cite{DeLaurentis:2022nrv}. The modification of the dynamical friction force due to the ULDM self-interaction was considered in \cite{Boudon:2022dxi,Hartman,Buehler:2022tmr,Berezhiani:2023vlo}, where it was shown that although this interaction is very weak, the corresponding modification could play an important role for the stellar motion \cite{Glennon}.   {Note that self-interaction is important not only for galactic ULDM BEC but also in the case of conventional condensed matter systems \cite{PhysRevB.88.184503,PhysRevA.89.033626}.}

In addition to stars, whose treatment as pointless objects is an excellent approximation, dynamical friction force should affect also extended moving bodies. One particularly important class of such bodies are globular clusters. They are large and dense agglomerates of stars with typical half-radius 3-5 pc up to tens of pc with masses $10^3M_\odot-10^6 M_\odot$  \cite{Gratton} and it is estimated that the Milky Way contains more than 150 globular clusters. Another important class of extended objects for the study of dynamical friction acting on moving extended objects are dwarf galaxies. Since dwarf galaxies typically move in halo of massive galaxies which is characterized by a different state of ultralight dark matter  {with nonzero temperature \cite{Chavanis:2018pkx}} compared to the core region, we leave the investigation of the dynamical friction force acting on moving dwarf galaxies for future studies.

Since globular clusters are much more massive than ordinary stars and the dynamical friction force is proportional to the square of the mass of a moving object, dynamical friction can have much stronger effect on their orbits compared to the case of stars \cite{Tremaine:1975}. Indeed, the study of the orbital motion of globular clusters in the Fornax dwarf spheroidal, which is a satellite of the Milky Way and contains six globular clusters \cite{Pace}, showed that the
decay timescale for these clusters to sink toward the galactic nucleus due to dynamical friction is $\sim 1$ Gyr \cite{Tremaine:1976}. The Fornax timing problem was considered in the ULDM model in \cite{Blas}. According to the study in Ref.\cite{Hartman}, while a zero-temperature ULDM superfluid yields decay times for globular clusters in Fornax which are in agreement with observations, finite temperature ULDM models may lead to very small decay times.  {It should be noted also that the dynamical friction force may provide an additional drag force on the cluster`s stars and potentially increase the cluster stability affecting its gravitational heating rate \cite{Marsh:2018zyw}.}

Globular clusters are not pointlike objects and are often modelled as Plummer spheres \cite{Plummer}.  {We would like to mention that the dynamical friction, in this case, is not simply the sum of the individual frictional forces from each star in a cluster because the de Broglie wavelength of the DM particle is much larger than the interstellar distances in a globular cluster.}  Numerical studies in \cite{Lancaster_2020,Glennon} showed that the dynamical friction force is weaker for the Plummer sphere compared to a point object of the same mass alleviating the timing problem. This result and the fact that globular clusters are spatially extended objects provides motivation for our study in the present paper, where we aim and derive an analytical formula for the dynamical friction force acting on circularly moving Plummer spheres in self-interacting ULDM environment.

The paper is organized as follows. The dynamical friction force acting on circularly moving Plummer spheres in the ULDM core region is considered in  Sec.\ref{sec:extended-body}. The radial and tangential components of the dynamical friction force are calculated and some illustrative numerical results are obtained in Sec.\ref{sec:dynamical-friction-components}. Conclusions are drawn in Sec.\ref{sec:Conclusion}.

\section{Total dynamical friction force and torque for circularly moving Plummer sphere}
\label{sec:extended-body}

In this section, we determine the total dynamical friction force and torque acting on a Plummer sphere, which moves on a circular orbit of radius $r_0$ with constant angular velocity $\Omega$ in the steady-state regime.

\begin{figure}[t]
  \begin{center}
    \includegraphics[width=0.47\textwidth]{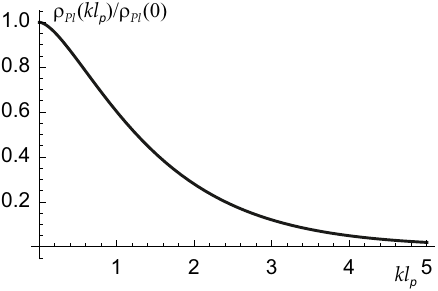}
    \caption{The Plummer sphere mass density in momentum space normalized by its value at zero momentum.}
     \label{Fig1:density}
    \end{center} 
\end{figure}

The Plummer sphere of radius $l_p$ and total mass $M$, whose mass density profile  {is spherical symmetric, depends on the radial coordinate $r$, and} is given by
\begin{equation}
\rho_{Pl}(\mathbf{r})=\frac{3M}{4\pi l^3_p}\frac{1}{\left(1+\frac{r^2}{l^2_p}\right)^{5/2}},
\label{Plummer-sphere}
\end{equation}
produces the gravitational potential which is often used in Monte Carlo simulations of globular clusters $U_{Pl}(r)=-\frac{GM}{\sqrt{r^2+l^2_p}}$, where $G$ is Newton`s gravitational constant. Using
 \cite{Gradshteyn}, we find the following Fourier transform of the Plummer sphere mass density:
$$
\rho_{Pl}(\mathbf{k})=
\frac{3M}{2l^3_p}\int^{\infty}_0 r^2dr\int^{\pi}_0 d\theta \sin\theta e^{-ikr\cos\theta}\frac{1}{\left(1+\frac{r^2}{l^2_p}\right)^{5/2}}=\frac{3Ml^2_p}{2ik}\int^{\infty}_{-\infty} \frac{xdx\,e^{ikx}}{\left(l^2_p+x^2\right)^{5/2}}=
$$
\begin{equation}
-\frac{Ml^2_p}{2ik}\int^{\infty}_{-\infty} e^{ikx}d\left(\frac{1}{\left(l^2_p+x^2\right)^{3/2}}\right)
=M\int^{\infty}_0 dx\frac{\cos(kl_px)}{\left(1+x^2\right)^{3/2}}=Mkl_p\, K_1(kl_p)\equiv \rho_{Pl}(kl_p),
\label{density-momentum-1}
\end{equation}
where $K_1(x)$ is the modified Bessel function of the second kind. Since the Fourier transform of the Plummer sphere mass density will play a crucial role in our subsequent analysis, we plot $\rho_{Pl}(kl_p)$ normalized by $\rho_{Pl}(0)$ in Fig.\ref{Fig1:density}.
For $l_p \to 0$, using the asymptote
$K_1(x) \to 1/x$ for $x \to 0$, we
find that $\rho_{Pl}(k l_p) \to M$,
i.e., the mass density profile of Plummer sphere in momentum space tends to the mass density profile of a point probe given by $\rho_p(\mathbf{k})=M$.

In our derivation of the dynamical friction force, we follow the setup developed in \cite{Desjacques_2022,Buehler:2022tmr}.
 {The granularity effects are important in analysing the dynamical friction force because orbits of globular clusters are usually situated in halos of massive galaxies which possess a granular structure \cite{Alvarez-Rios:2023cch}. However, to proceed with the analytic approach we consider in this paper the most simple situation of homogeneous ULDM density.
In this case,} the Plummer sphere moving in a homogeneous ultra-light dark matter composed of ultra-light bosonic particles of mass $m$ perturbs due to gravitational interaction the ULDM density $\rho_{DM}(t,\mathbf{r})=\rho_0(1+\alpha(t,\mathbf{r}))$.  {The linear response approach assumes that the change of ULDM density caused by a moving body is small. To test the validity of the linear response approach fully non-linear simulations were performed in \cite{Lancaster_2020}. Comparing these numerical simulations to analytic results, an excellent agreement with the linear regime was found}.

  {The dynamical friction force is determined by the gravitational potential induced by perturbation of the ULDM density by a moving body. It can be found by solving the system of the Euler and continuity equations for the ULDM density and velocity as well as taking into account the Poisson equation for the gravitational potential, see, for details \cite{Hui:2016ltb,Desjacques_2022}.}
 In the linear response approach, the corresponding density inhomogeneity $\alpha(t,\mathbf{r})$ is governed by the equation
\begin{equation}
\partial_t^2\alpha-c^2_s\nabla_{\mathbf{r}}^2\alpha+\frac{\nabla_{\mathbf{r}}^4\alpha}{4m^2}=4\pi G\rho_{Pl}(\mathbf{r}-\mathbf{r}_{CM}(t)),
\label{perturbation-sphere}
\end{equation}
where $\mathbf{r}_{CM}(t)$ denotes the position of the center of mass of the moving Plummer sphere and $c_s$ is the adiabatic sound velocity of DM superfluid. We neglected on the light-hand side of the above equation the self-gravity term $-4\pi G\rho_0\alpha$. This term is related to the Jeans instability and its role on the dynamical friction was considered in \cite{Elder}. It was shown that this term is responsible for small non-zero dynamical friction even for subsonic linear motion of perturber in superfluid, where Landau`s criterion of superfluidity forbids the existence of dynamical friction if the self-gravity term is absent. Note that density inhomogeneity in the case of point probe satisfies also Eq.(\ref{perturbation-sphere}) with replacement $\rho_{Pl}(\mathbf{r}-\mathbf{r}_{CM}(t)) \to M\delta^3(\mathbf{r}-\mathbf{r}_p(t))$, where $\mathbf{r}_p(t)$ describes the trajectory of a moving point probe \cite{Desjacques_2022,Gorkavenko_2024}. In momentum space, Eq.(\ref{perturbation-sphere}) takes the following form for the steady-state motion:
\begin{equation}
\left(-\omega^2+c^2_s\mathbf{k}^2+\frac{\mathbf{k}^4}{4m^2}\right)\alpha(\omega,\mathbf{k})=4\pi G \int^{+\infty}_{-\infty} d\tau\,e^{i\omega\tau-i\mathbf{k}\mathbf{r}_{CM}(\tau)}\rho_{Pl}(\mathbf{k}),
\label{perturbation-sphere-momentum-1}
\end{equation}
where $\rho_{Pl}(\mathbf{k})$ is the mass density profile of the Plummer sphere in momentum space.
Note that Eq.(\ref{perturbation-sphere-momentum-1}) differs from the corresponding equation for ULDM perturbation in the case of a point probe  {(see, e.g., \cite{Berezhiani:2023vlo})} only by the presence of factor $\rho_{Pl}(\mathbf{k})$ instead of $\rho_p(\mathbf{k})=M$ in the integrand on the right-hand side.

According to the Poisson equation, the density inhomogeneity $\alpha(t,\mathbf{r})$ and the Plummer sphere density $\rho_{Pl}(t,\mathbf{r})$ source a perturbation $\phi$ of the gravitational potential defined by the equation
\begin{equation}   {\vec\nabla}^2\phi(t,\mathbf{r})=4\pi G(\rho_0\alpha(t,\mathbf{r})+\rho_{Pl}(\mathbf{ r}-\mathbf{r}_{CM}(t))).
\label{gravitational-potential}
\end{equation}
Solving Eq.(\ref{perturbation-sphere-momentum-1}) for $\alpha(\omega,\mathbf{k})$, Eq.(\ref{gravitational-potential}) gives the following gravitational potential $\phi_{\alpha}$ due to perturbed ULDM density by the moving Plummer sphere (the complete perturbed gravitational potential is obviously $\phi=\phi_{\alpha}+\phi_{Pl}$, where $\phi_{Pl}$ is the Newtonian potential of the Plummer sphere):
\begin{equation}
\phi_{\alpha}(t,\mathbf{r})=-4\pi G\rho_0 \int \frac{d\omega d^3k}{(2\pi)^4}\frac{\alpha(\omega,\mathbf{k})}{\mathbf{k}^2}\,e^{-i\omega t+i\mathbf{k}\mathbf{r}}.
\end{equation}
Then the local dynamical friction force density acting on the moving Plummer sphere is given by
$$
\mathbf{f}_{fr}(t,\mathbf{r})=-\rho_{Pl}(\mathbf{r}-\mathbf{r}_{CM}(t))\nabla_{\mathbf{r}}\phi_{\alpha}(t,\mathbf{r})
=4\pi G\rho_0 \rho_{Pl}(\mathbf{r}-\mathbf{r}_{CM}(t))\int \frac{d\omega d^3k}{(2\pi)^4}\frac{i\mathbf{k}}{\mathbf{k}^2}\alpha(\omega,\mathbf{k})e^{-i\omega t+i\mathbf{k}\mathbf{r}}
$$
\begin{equation}
=(4\pi G)^2\rho_0 \rho_{Pl}(\mathbf{r}-\mathbf{r}_{CM}(t))\int^{+\infty}_{-\infty}d\tau\int \frac{d\omega d^3k}{(2\pi)^4}\frac{i\mathbf{k}}{\mathbf{k^2}}\frac{e^{-i\omega (t-\tau)+i\mathbf{k}\mathbf{r}-i\mathbf{k}\mathbf{r}_{CM}(\tau)}\rho_{Pl}(\mathbf{k})}{-(\omega+i\epsilon)^2+c^2_s\mathbf{k}^2+\frac{\mathbf{k}^4}{4m^2}},
\label{dynamical-force-local}
\end{equation}
where $\epsilon \to +0$ takes into account the retarded character of ULDM perturbation by the moving Plummer sphere. It is convenient to change the variable $\tau=-\tau^{\prime} +t$ and then take into account that the integral over $\omega$ vanishes for $\tau^{\prime}<0$. We get
\begin{equation}
\mathbf{f}_{fr}(t,\mathbf{r})=(4\pi G)^2\rho_0 \rho_{Pl}(\mathbf{r}-\mathbf{r}_{CM}(t))\int^{+\infty}_{0}d\tau^{\prime} \int \frac{d\omega d^3k}{(2\pi)^4}\frac{i\mathbf{k}}{\mathbf{k^2}}\frac{e^{-i\omega\tau^{\prime}+i\mathbf{k}\mathbf{r}-i\mathbf{k}\mathbf{r}_{CM}(t-\tau^{\prime})}\rho_{Pl}(\mathbf{k})}{-(\omega+i\epsilon)^2+c^2_s\mathbf{k}^2+\frac{\mathbf{k}^4}{4m^2}}.
\label{dynamical-force-local-1}
\end{equation}
Clearly, this local dynamical force is different at different position $\mathbf{r}$ within the Plummer sphere producing tidal force and torque also. We will not study tidal force in this paper, however, will discuss briefly torque below.

Since the most important characteristics of dynamical friction is the total dynamical friction force acting on moving extended body, we integrate the local dynamical friction force (\ref{dynamical-force-local}) over $\mathbf{r}$ denoting for convenience $\tau^{\prime}$ by $\tau$. We obtain the following total dynamical friction force:
\begin{equation}
\mathbf{F}_{fr}(t)=(4\pi G)^2\rho_0\int^{+\infty}_{0}\!\!\! d\tau\int d^3r \rho_{Pl}(\mathbf{r}-\mathbf{r}_{CM}(t)) \int \frac{d\omega d^3k}{(2\pi)^4}\,\frac{i\mathbf{k}}{\mathbf{k}^2}\,\frac{\rho_{Pl}(\mathbf{k})\,e^{-i\omega\tau+i\mathbf{k}\mathbf{r}-i\mathbf{k}\mathbf{r}_{CM}(t-\tau)}}{-(\omega+i\epsilon)^2+c^2_s\mathbf{k}^2+\frac{\mathbf{k}^4}{4m^2}}.
\label{dynamical-force-sphere-total0}
\end{equation}
Since the density profile of the moving Plummer sphere in the expression above depends in coordinate space actually on $\mathbf{r}-\mathbf{r}_{CM}(t)$, it is convenient to make the change of variable $\mathbf{r}=\mathbf{r}^{\prime}+\mathbf{r}_{CM}(t)$. Then using $e^{i\mathbf{k}\mathbf{r}}=e^{i\mathbf{k}\mathbf{r}^{\prime}+i\mathbf{k}\mathbf{r}_{CM}(t)}$ and integrating over $\mathbf{r}^{\prime}$, we find
the following expression for the total dynamical friction force acting on the moving Plummer sphere:
\begin{equation}
\mathbf{F}_{fr}(t)=(4\pi G)^2\rho_0\int^{+\infty}_{0} d\tau\int \frac{d\omega d^3k}{(2\pi)^4}\,\frac{i\mathbf{k}}{\mathbf{k}^2}\,\frac{\rho_{Pl}(-\mathbf{k})\rho_{Pl}(\mathbf{k})\,e^{-i\omega\tau+i\mathbf{k}\mathbf{r}_{CM}(t)-i\mathbf{k}\mathbf{r}_{CM}(t-\tau)}}{-(\omega+i\epsilon)^2+c^2_s\mathbf{k}^2+\frac{\mathbf{k}^4}{4m^2}}.
\label{dynamical-force-sphere-total}
\end{equation}

As to the total torque due to the dynamical friction force, using Eq.(\ref{dynamical-force-local-1}), we find that it equals\vspace{-0.55em}
$$
 \mbox{\boldmath $\tau$} = \int d^3r\,\mathbf{r}\times\mathbf{f}_{fr}(t,\mathbf{r}) \equiv \int d^3r\,(\mathbf{r}-\mathbf{r}_{CM}(t))\times\mathbf{f}_{fr}(t,\mathbf{r})+\mathbf{r}_{CM}(t)\times\mathbf{F}_{fr}(t),
$$
where the last term $\mathbf{r}_{CM}(t)\times\mathbf{F}_{fr}(t)$ is torque due to the total dynamical friction force which describes the loss of the angular moment of the Plummer sphere in its orbital motion. The other term in the above equation defines torque $\tau_{inner}$ which does not have an analog in the case of a point probe. It may cause rotation of the Plummer sphere as a whole. This torque equals
$$
\mbox{\boldmath $\tau$}_{inner}(t)
=(4\pi G)^2\rho_0\int d^3r\,\rho_{Pl}(\mathbf{r}-\mathbf{r}_{CM}(t))\int^{+\infty}_{0}d\tau
$$
\begin{equation}
\times \int \frac{d\omega d^3k\,e^{-i\omega\tau+i\mathbf{k}\mathbf{r}-i\mathbf{k}\mathbf{r}_{CM}(t-\tau)}}{(2\pi)^4}\frac{i(\mathbf{r}-\mathbf{r}_{CM}(t))\times\mathbf{k}}{\mathbf{k^2}}\frac{\rho_{Pl}(\mathbf{k})}{-(\omega+i\epsilon)^2+c^2_s\mathbf{k}^2+\frac{\mathbf{k}^4}{4m^2}}.
\label{torque-local}
\end{equation}
Making the change of variable $\mathbf{r}=\mathbf{r}^{\prime}+\mathbf{r}_{CM}(t)$, we obtain
$$
\mbox{\boldmath $\tau$}_{inner}(t)
=(4\pi G)^2\rho_0\int^{+\infty}_{0}d\tau \int \frac{d\omega d^3k\,e^{-i\omega\tau+i\mathbf{k}\mathbf{r}_{CM}(t)-i\mathbf{k}\mathbf{r}_{CM}(t-\tau)}}{(2\pi)^4}
$$
\begin{equation}
\times \int d^3r^{\prime}\,e^{i\mathbf{k}\mathbf{r}^{\prime}}\rho_{Pl}(\mathbf{r}^{\prime})\frac{i\mathbf{r}^{\prime}\times\mathbf{k}}{\mathbf{k^2}}\frac{\rho_{Pl}(\mathbf{k})}{-(\omega+i\epsilon)^2+c^2_s\mathbf{k}^2+\frac{\mathbf{k}^4}{4m^2}}.
\label{torque-local-1}
\end{equation}
Obviously, torque $\mbox{\boldmath $\tau$}_{inner}$ vanishes in the case of Plummer sphere, which is spherically symmetric, because
\begin{equation}\label{gradk}
\int d^3r^{\prime}\,e^{i\mathbf{k}\mathbf{r}^{\prime}}\rho_{Pl}(\mathbf{r}^{\prime})\,i\mathbf{r}^{\prime}=\nabla_{\mathbf{k}}\int d^3r^{\prime}\,e^{i\mathbf{k}\mathbf{r}^{\prime}}\rho_{Pl}(\mathbf{r}^{\prime})=\nabla_{\mathbf{k}}\rho_{Pl}(\mathbf{k})
\end{equation}
is proportional to vector $\mathbf{k}$ taking into account that $\rho_{Pl}(\mathbf{k})$ depends only on $k=|\mathbf{k}|$ and its vector product with $\mathbf{k}$ in Eq.(\ref{torque-local-1}) gives zero, i.e., the dynamical friction does not produce rotation of the Plummer sphere as a whole.
As it is clear from Eq.\eqref{gradk} the inner torque $\mbox{\boldmath $\tau$}_{inner}$ may not vanish for a non-spherically symmetric body, e.g., an extended body deformed by tidal forces. 


\section{Radial and tangential components of dynamical friction force}
\label{sec:dynamical-friction-components}

Let us determine the radial and tangential components of the total dynamical friction force acting on the Plummer sphere, which moves on a circular orbit of radius $r_0$ with constant angular velocity $\Omega$. Hence, the orbital velocity of the Plummer sphere is $v=\Omega r_0$.

The total dynamical friction force (\ref{dynamical-force-sphere-total}) differs from the expression for the dynamical friction force acting on the moving point probe given by Eq.(33) in \cite{Berezhiani:2023vlo} only by the presence of the product of the density profiles in momentum space divided by $M^2$, i.e., $\rho_{Pl}(\mathbf{k})\rho_{Pl}(-\mathbf{k})/M^2$. Since $\rho_{Pl}(\mathbf{k})$ depends only on the absolute value of momentum $k$ for the case of a spherically symmetric body considered in the present paper, the subsequent integration $d^3k=k^2dkd\Omega_k$ over $\Omega_k$ proceeds as in \cite{Berezhiani:2023vlo} and we obtain the following total dynamical friction force:
\begin{equation}
\mathbf{F}_{fr}(t)=-\frac{4\pi G^2M^2\rho_0}{c^2_s}\vec {\mathcal{F}},
\label{dynamical-force}
\end{equation}
where $\vec {\mathcal{F}}$ is dimensionless force whose nonzero radial and tangential components are given by
\begin{equation}
\vec {\mathcal{F}}=\sum_{\ell=1}^{\ell_\text{\tiny max}}\sum_{m_l=-\ell}^{\ell-2}\gamma_{\ell m_l}\left\{\text{Re}\left(S_{\ell,\ell-1}^{m_l}-{S^{m_l+1}_{\ell,\ell-1}}^*\right)\hat{r}+\text{Im}\left(S_{\ell,\ell-1}^{m_l}-{S^{m_l+1}_{\ell,\ell-1}}^*\right)\hat{\varphi}\right\}.
\label{FDF1}
\end{equation}
Here
\begin{multline}
    \gamma_{\ell m_l}=  (-1)^{m_l} \frac{(\ell-m_l)!}{(\ell-m_l-2)!}\\ \times\left\{{\Gamma\left(\frac{1-\ell-m_l}{2}\right)\Gamma\left(1+\frac{\ell-m_l}{2}\right)\Gamma\left(\frac{3-\ell+m_l}{2}\right)\Gamma\left(1+\frac{\ell+m_l}{2}\right)}\right\}^{-1}\,,
\end{multline}
and the key quantity which defines the dynamical friction force is
\begin{equation}
S^{m_l}_{\ell,\ell-1}
=\frac{c^2_s}{M^2}\int^{+\infty}_0 \frac{kdk\,\rho^2_{Pl}(k l_p)\,j_{\ell}(kr_0)j_{\ell-1}(kr_0)}{c^2_sk^2+\frac{k^4}{4m^2}-(m_l\Omega+i\epsilon)^2},\quad \epsilon \to +0,
\label{S-function-1}
\end{equation}
where $\ell$ and $m_l$ are the azimuthal and quantum numbers, respectively, $j_{\ell}(x)$ is the spherical Bessel function. As was pointed out in \cite{Berezhiani:2023vlo} the probe size acts as the effective cutoff $l_{max}$ in the sum over $l$ given by
$$
\ell_\text{\tiny max}=\frac{\pi r_0}{l_p}.
$$

Obviously, $S^{m_l}_{\ell,\ell-1}$ is real for $m_l=0$, therefore, $S^0_{\ell,\ell-1}$ does not contribute to the tangential component of the dynamical friction force, which is defined by the imaginary part of $S^{m_l}_{\ell,\ell-1}$. For $m_l \ne 0$, it is not difficult to calculate the imaginary part of $S^{m_l}_{\ell,\ell-1}$ by applying the Sokhotski-Plemelj formula
$$
\frac{1}{f(x)\pm i\epsilon}=p.v. \left(\frac{1}{f(x)}\right)\mp \pi i\delta(f(x)),\quad \epsilon \to +0
$$
to Eq.(\ref{S-function-1}). We find
\begin{equation}
Im\,S^{m_l}_{l,l-1}=\frac{4m^2c^2_s\pi}{M^2}\mbox{sgn}(m_l)\int^{+\infty}_0 kdk\,\rho^2_{Pl}(k l_p)\,j_{l}(kr_0)j_{l-1}(kr_0)\,\delta(4m^2c^2_sk^2+k^4-4m^2m^2_l\Omega^2).
\label{S-function-Plummer}
\end{equation}

Representing the argument of the $\delta$-function as
$$
4m^2c^2_sk^2+k^4-4m^2m^2_l\Omega^2=
(k-k_1)(k-k_2)(k-k_3)(k-k_4),
$$
where
\begin{equation}
k_{1,2}=\pm mc_sif^+_{m_l},\quad k_{3,4}=\pm mc_sf^-_{m_l}
\label{poles}
\end{equation}
with
$$
f^+_{m_l}=\sqrt{2}\left(\sqrt{1+\frac{m_l^2\Omega^2}{m^2c^4_s}}+1\right)^{1/2},\quad f^-_{m_l}=\sqrt{2}\left(\sqrt{1+\frac{m_l^2\Omega^2}{m^2c^4_s}}-1\right)^{1/2}
$$
define zeros of the argument, we can easily calculate the integral over $k$ because only $k_3$ is real and positive\footnote{For the numerical computation one has to restore the Planck constant $\hbar$ in expressions for $k_i$ and $f^\pm_{m_l}$, i.e. $k_i\rightarrow k_i/\hbar$ and $\Omega\rightarrow \hbar \Omega$.}. We obtain
\begin{equation}
Im\,S^{m_l}_{\ell,\ell-1}
=\pi\,\mbox{sgn}(m_l)\,\frac{\rho^2_{Pl}(k_3 l_p)}{M^2}\frac{j_{\ell}(k_3r_0)j_{\ell-1}(k_3r_0)}{2\sqrt{1+\frac{\hbar^2\Omega^2}{m^2 c^4_s }\,m_l^2 }},
\label{S-function-Plummer-1}
\end{equation}
where $\rho_{Pl}(k l_p)$ is the mass density of Plummer sphere given by Eq.\eqref{density-momentum-1}.
Note that setting $ \rho_{Pl}(k_3 l_p)=M$ (for $l_p\rightarrow 0$) the above equation reproduces exactly the imaginary part of $S^{m_l}_{l,l-1}$ for a point probe found in \cite{Berezhiani:2023vlo}.

We calculate numerically the real part of $S^{m_l}_{\ell,\ell-1}$ given by the Cauchy principal value of the integral in Eq.(\ref{S-function-1}) 
\begin{equation}
Re\, S^{m_l}_{\ell,\ell-1}=
{\rm p.v.}\,\frac{4 m^2 c^2_s r_0^2}{ \hbar^2M^2} \int^{+\infty}_0 \frac{x dx\,\rho^2_{Pl}(x l_p/r_0)\,j_{\ell}(x)j_{\ell-1}(x)}{x^4+\frac{4m^2c^2_sr_0^2}{\hbar^2}\, x^2-4 m_l^2\,\frac{ m^2\Omega^2r_0^4}{\hbar^2} }.
\label{S-function-1vp}
\end{equation}
Since  $\rho_{Pl}(x)\to M$ for $x\to 0$, our relation reproduces exactly the real part of $S^{m_l}_{l,l-1}$ for a point probe when $l_p\rightarrow0$.

\begin{figure}[t!]
  \begin{center}
    \includegraphics[width=0.95\textwidth]{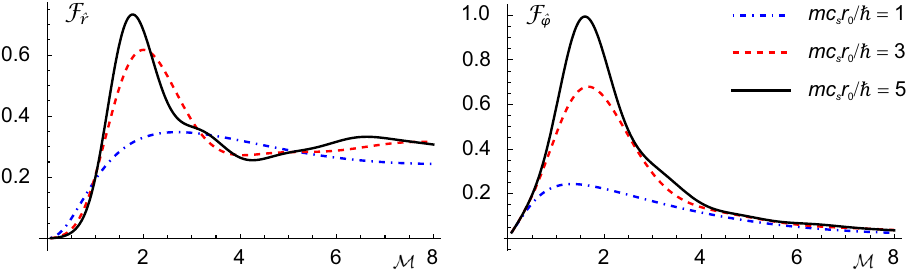}
    \caption{Radial (left panel) and tangential (right panel) components of the dimensionless dynamical friction force $\vec {\mathcal{F}}$ for a point probe as a function of the Mach number $\mathcal{M}$ for a few values of orbital radius $r_0$.}
     \label{Fig1}
    \end{center} 
\end{figure}

Let us begin our analysis by considering the behaviour of the radial and tangential components of the dimensionless dynamical friction force $\vec {\mathcal{F}}$ for a point probe for different values of orbital radius $r_0$ as a function of the Mach number $\mathcal{M}=v/c_s$. Setting $l_p=0$ in Eqs.(\ref{S-function-Plummer-1}) and (\ref{S-function-1vp}) and using Eq.(\ref{FDF1}), we show our numerical results for the radial and tangential components of the dynamical friction force in Fig.\ref{Fig1}. Clearly, both components have a maximum at a certain Mach number. The position of the maximal value of the radial component shifts to smaller values of $\mathcal{M}$ as $r_0$ grows. On the other hand, the position of the maximal value of the tangential component shifts to larger values of $\mathcal{M}$ as $r_0$ increases. For $\mathcal{M}\rightarrow 0$, both components of the dynamical friction force tend to zero.

\begin{figure}[t]
  \begin{center}
    \includegraphics[width=0.95\textwidth]{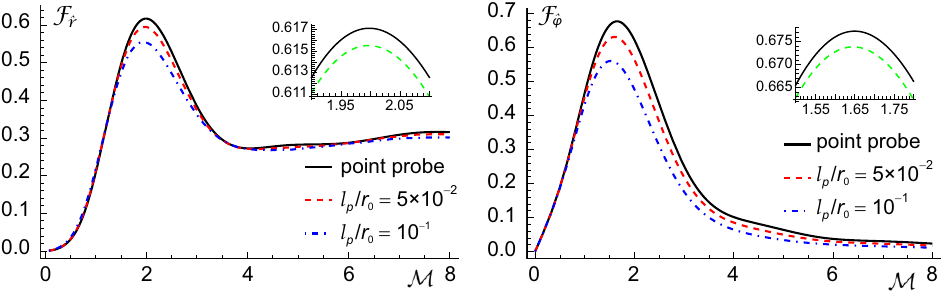}
    \caption{Radial (left panel) and tangential (right panel) components of the dimensionless dynamical friction force $\vec {\mathcal{F}}$  as a function of the Mach number $\mathcal{M}$ at fixed orbital radius $r_0 = 3\hbar/(m c_s)$ for point probe and the Plummer sphere. 
    Insets show the radial and tangential components of the dimensionless dynamical friction force in the vicinity of its maximum for a point probe (black solid line) and Plummer sphere with $l_p/r_0=10^{-2}$ (green dashed line).
}
     \label{Fig2}
    \end{center} 
\end{figure}

Using Eqs.\eqref{FDF1}, \eqref{S-function-Plummer-1}, and \eqref{S-function-1vp} we determine numerically the radial and tangential components of the dimensionless dynamical friction force $\vec {\mathcal{F}}$ acting on the circularly moving Plummer sphere and plot them as a function of the Mach number ${\cal M}$ for $l_p/r_0=5\cdot 10^{-2}$ and $l_p/r_0=10^{-1}$ in Fig.\ref{Fig2}. To compare with results for a point probe of the same mass we fix the value of the orbital radius $r_0 = 3\hbar/(m c_s)$ considered in \cite{Berezhiani:2023vlo}. Since the dynamical friction forces are practically the same for the Plummer sphere and point probe for the ratio smaller than $l_p/r_0\sim 10^{-2}$, to show their difference we plot in insets the radial and tangential components of the dimensionless dynamical friction force for $l_p/r_0=10^{-2}$ in the vicinity of its maximum.  Our results show that the dynamical friction force is practically the same for the Plummer sphere and point probe at small values of the Mach number $\mathcal{M}$. This force differs most strongly at its maximum.

\begin{figure}[b!]
  \begin{center}
    \includegraphics[width=0.95\textwidth]{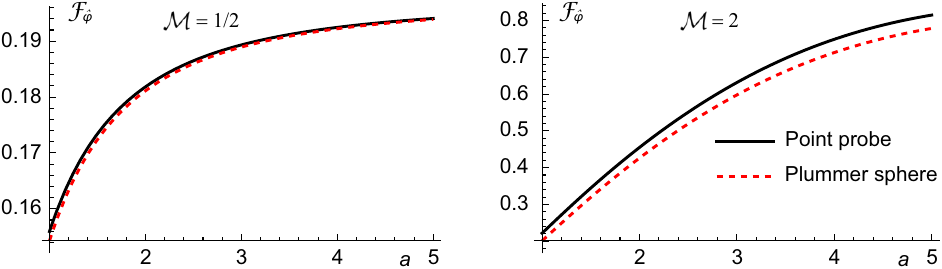}
    \caption{The tangential component of the dimensionless dynamical friction force for the Mach number ${\mathcal{M}}=1/2$ (left panel) and ${\mathcal{M}}=2$ (right panel) as a function of dimensionless orbital radius $a =m c_sr_0/ \hbar$ at fixed $l_p=\hbar/(10 m c_s)$ for point probe and the Plummer sphere.}
     \label{Fig3}
    \end{center} 
\end{figure}

\begin{figure}[t]
  \begin{center}
    \includegraphics[width=0.95\textwidth]{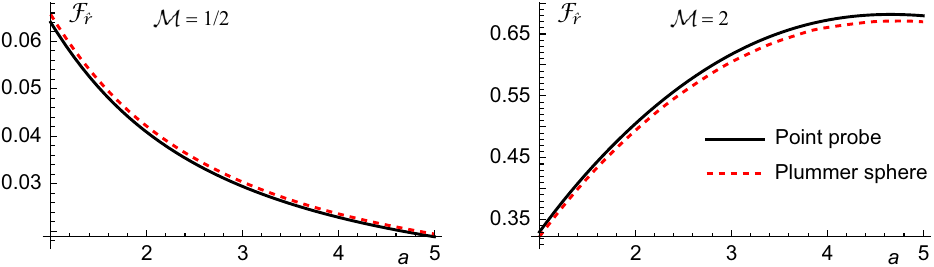}
    \caption{The radial component of the dimensionless dynamical friction force for the Mach number ${\mathcal{M}}=1/2$ (left panel) and ${\mathcal{M}}=2$ (right panel) as a function of dimensionless orbital radius $a =m c_sr_0/ \hbar$ at fixed $l_p=\hbar/(10 m c_s)$ for point probe and the Plummer sphere.}
     \label{Fig4}
    \end{center} 
\end{figure}

Since the dynamical friction force depends on two parameters $\mathcal{M}$ and $r_0$, to display the dependence on the latter parameter we plot the radial and tangential components of the dimensionless dynamical friction force $\vec {\mathcal{F}}$ in Figs.\ref{Fig3} and \ref{Fig4} as a function of dimensionless orbital radius $a =m c_sr_0/ \hbar$ in the interval $a\in [1,5]$.
Two cases of a point probe and the Plummer sphere with fixed radius $l_p=\hbar/(10 m c_s)$ are considered for the Mach number $\mathcal M=0.5$ and $\mathcal M=2$. 
This dependence is qualitatively different for the radial and tangential components. While the tangential component monotonously grows, the radial component decreases with $a$ for $\mathcal M=0.5$, but increases for $\mathcal M=2$.

\section{Conclusions}
\label{sec:Conclusion}

Globular clusters can be more correctly described as extended rather than point objects and are often modelled as Plummer spheres in numerical calculations. In this paper, we investigated how the finite size of the object affects the dynamic friction force. We derived an analytic formula for the total dynamical friction force acting on the Plummer sphere moving on a circular orbit with constant angular velocity in ultralight bosonic dark matter. Both radial and tangential components of the dynamical friction force were determined.

 {The model we used has some limitations. First of all, our results are given for ULDM particles in a certain mass range when the de Broglie wavelength is significantly larger than the characteristic size of a globular cluster.
Then we used a spherically symmetric Plummer model to describe globular clusters ignoring the deformation of globular clusters due to tidal forces. In addition, we assumed that ULDM density is homogeneous and neglected its granular structure in halos of massive galaxies.
Thus, to overcome the limitations of our model, one has to take into account the tidal deformation of globular clusters, solve the Schroedinger-Poisson equation beyond the linear response approach, and account for the ULDM granular structure of the halo of massive galaxies.}


We compared the dynamical friction force for the Plummer sphere and point object of the same mass. For globular clusters and realistic values of the ratio of the Plummer sphere radius and the orbital radius, the finite size of the Plummer sphere practically does not affect the dynamical friction force if the ratio of the Plummer sphere radius to its orbital radius is less than approximately $10^{-2}$. 
Since the dynamical friction force $\vec {\mathcal{F}}$ depends on two parameters $\mathcal{M}$ and $r_0$, we determined the dependence of the radial and tangential components of $\vec {\mathcal{F}}$ on these parameters, shown in Figs.\ref{Fig2}, \ref{Fig3}, and \ref{Fig4}.

 {We would like to mention also that observationally, dynamical friction in ULDM halos can impact the orbital evolution of globular clusters and dwarf galaxies, potentially leading to measurable orbital decay. Precise tracking of these orbital changes, especially in Milky Way satellites like Fornax, can set constraints on the mass of bosonic particles and self-interaction strength. Comparing theoretical models of orbital decay with observations could help infer the self-interaction term in the Lagrangian, offering a potential method to distinguish between weakly and strongly interacting ULDM scenarios.}

For simplicity, we neglected tidal forces and assumed complete spherical symmetry. Certainly, the question how tidal forces and torque for spherically asymmetric extended bodies affect dynamical friction is definitely interesting and deserves an in depth study. In addition, it would be interesting to extend the present study to dwarf galaxies moving in halos of massive galaxies, where the ULDM thermal pressure dominates over the core region quantum pressure considered here.


\vspace{5mm}

\centerline{\bf Acknowledgements}
\vspace{5mm}

A.I.Ya. acknowledges support from the PRIN Project ‘Quantum Atomic Mixtures: Droplets, Topological Structures, and Vortices’. 
The work of E.V.G. and V.M.G. was partially supported by the project
’Fundamental laws of physics in cosmology of the early Universe’ of the Ministry of Education and Science of Ukraine (22BF051-06).


\bibliographystyle{JHEP}
\bibliography{bibliography}

\end{document}